\documentclass[12pt,draft]{IEEEtran}
\usepackage{amsmath,amsfonts,amssymb}
\usepackage{psfig}
\usepackage{pst-plot}
\usepackage{latexsym}
\usepackage{hhline,eufrak,euscript,delarray}
\usepackage[ruled,vlined]{algorithm2e}   % ALGORITHM PACK
\newtheorem{theorem}{Theorem}
\newtheorem{lemma}[theorem]{Lemma}
\newtheorem{corollary}[theorem]{Corollary}

\newtheorem{thm}[theorem]{Theorem}

\def \beq{\begin{equation}}
\def \eeq{\end{equation}}
\def \be{\begin{eqnarray*}}
\def \ee{\end{eqnarray*}}
\def \ben{\begin{eqnarray}}
\def \een{\end{eqnarray}}

\def \L{\left}
\def \R{\right}
\def \E{{\mathbb{R}}} % Real Euclidian Space

\def \hup{N^{+}}
\def \hdown{N^{-}}
\def \Pr{{\mathbb{P}}}  % PROBA
\def \Esp{{\mathbb{E}}} % MEAN

\def\coeff#1{\left[ #1 \right]}
\def\ten{\rightarrow}
\def\erfc{\mbox{erfc}}
\def\DIAM{D}

\def\jmin{j}

\def\REGION{\mathcal{R}}
\def\SREGION{\mathcal{S}}
\def\VOLUME{{V}}
\def\RRR{\mbox{\textsc{r}}\normalsize}
\def\SENSE{{\RRR}_{\tiny\mbox{\textsc{Sense}}\normalsize}}
\def\TRANS{{\RRR}_{\tiny\mbox{\textsc{Trans}}\normalsize}}

\def\GPOISSON{{\mathbb{G}}^{\mbox{\tiny Poisson \normalsize}}(\rho, \,\RRR(\rho))}
\def\GPX{{\mathbb{G}}_{\REGION}^{\mbox{\tiny Poisson \normalsize}}(\rho, \,\RRR(\rho))}
\def\G{{\mathbb{G}}}

\begin{document}
\title{Extremal Properties of Three Dimensional Sensor Networks %
with Applications}
\author{Vlady Ravelomanana\thanks{Vlady Ravelomanana is with the
LIPN -- CNRS UMR 7030, Institut Galil\'ee
 Universit\'e de Paris 13, France. 
 E-mail~: vlad@lipn.univ-paris13.fr}}
%% URL~: http://www-lipn.univ-paris13.fr/~ravelomanana}
\maketitle

\noindent
\begin{abstract}
In this paper, we analyze various critical transmitting/sensing
ranges for connectivity and coverage in three-dimensional sensor networks.
As in other large-scale 
complex systems, many global parameters of sensor
networks undergo phase transitions: For a given property of the network,
there is a critical threshold, corresponding to the minimum amount of 
the communication effort or
 power expenditure by individual nodes, above (resp. below) which the property exists with
high (resp. a low) probability. For sensor networks, properties of interest 
 include simple and multiple  degrees of connectivity/coverage.
First, we investigate the network topology according to the region of deployment, 
the number of deployed
sensors and their transmitting/sensing ranges.
More specifically, we consider the following problems:
Assume that $n$ nodes, each capable of sensing events
within a radius of $r$, are randomly and uniformly distributed
in a $3$-dimensional region $\REGION$ of volume $\VOLUME$, 
how large must the sensing range
$\SENSE$ be to ensure a given degree of coverage
  of the region to monitor? For a given transmission range $\TRANS$, what is the
minimum (resp. maximum) degree of the network?
What is then the typical hop-diameter of the underlying network?
Next, we show how these results affect algorithmic 
aspects of the network by designing
 specific distributed protocols for sensor networks.
\end{abstract}


%\end{frontmatter}

\section{Introduction}
Advances in micro-electro-mechanical systems technology,
wireless communications and digital electronics have enabled
the development of multi-functional sensor nodes. 
Sensor nodes are small miniaturized devices which 
 consist of sensing, data processing and
communicating components \cite{SURVEY,CHALLENGES,PERKINS}.
These inexpensive tiny sensors can be embedded or scattered
 onto target environments in order to monitor useful informations
in many situations. We categorize their applications into
law enforcement, environment, health, home and other commercial
areas. Moreover, it is possible to expand this classification
 with more categories including space exploration \cite{SPACE}
 and undersea monitoring \cite{WATER}. We refer here to the
survey paper \cite[Section 2]{SURVEY} for an extensive list of possible
applications of sensor networks. These critical applications introduce
the fundamental requirement of \textit{sensing coverage} that does not
exist in traditional ad hoc networks. In a \textit{sensing-covered}
network, every point of a targeted geographic region must be within
the sensing range of at least one sensor.

In general, networked sensors are very large systems, comprised of a vast
number of homogeneous miniaturized devices that cooperate
to achieve a sensing task. Each node
is equipped with a set of sensing monitors (light, pressure,
humidity, temperature, etc.) accordingly  to their designated tasks 
 and uses radio transmitters in order to communicate. 
 We refer, for example, to the web-sites \cite{BERKELEY,UCLA,JPL,WINS}
 of some  research institutions dedicated to the study and development of these networks.
 Typically, a sensor node is able to sense events within a given radius (the
 sensing range). Similarly, any pair of sensors are able to communicate 
 if they are within a distance less than their \textit{transmitting range}
  of each other. Wireless sensor networks are usually multihop
networks as opposed to wireless LAN environments.

A commonly encountered model of sensor network is defined
by a pair $n$ and $\REGION$ where $n$ homogeneous sensor nodes  
are randomly thrown in
a given region $\REGION$ of volume $\VOLUME=|{\REGION}|$, 
uniformly and independently.
This typical modeling assumption is commonly used by many researchers
\cite{BETTSTETTER,GRECS,CHENG,GILBERT,Kumar-Gupta,SKY,%
KRISHNA3,PHILIPS-PANWAR-TANTAWI,BLOUGH,SHAKKOTTAI}. In particular, the initial
 placement of the nodes is assumed to be random 
when the sensors are distributed
 over a region from a moving vehicle such as an airplane.
As opposed to traditional ad hoc networks, a sensor 
network is normally composed
 of nodes whose number can be several orders of magnitude higher than
the nodes in an ad hoc network. These sensor nodes are often deployed
 inside a phenomenon. Therefore, the positions of the nodes need not
be engineered or pre-determined.
Note that many existing results focused on planar networks 
\cite{GRECS,MCDIARMID,PHILIPS-PANWAR-TANTAWI}
 while three-dimensional settings reflect more accurately 
real-life situations \cite{SKY,SPACE}. Three-dimensional networks arise 
for instance in building networks where nodes are located in different floors.

 The aim of this paper is two-fold: (a) to study 
 the role that randomness plays in sensor networks
 and (b) to investigate the design and analysis of 
appropriate protocols for these networks.
These issues are motivated by the following simple reasons.
On first hand, the random placement of the nodes
 allows rapid deployment in
inaccessible terrains. On the other hand, this implies that the network
 must have self-organizing capabilities. 
Therefore, we start deriving analytical expressions
to characterize the topological properties of sensor networks.
Next, we discuss how 
to use these fundamental characteristics in order to design
and analyze fundamental algorithms such as those
  arising frequently in classical distributed systems.
To name a few, these algorithms include
 the broadcasting and gossiping protocols \cite{EXP-GAP,BASAGNI}, 
the leader election algorithm \cite{LEADER} and 
distributed code assignments protocols \cite{COLORS}. 

\section{Related  work}
Study of random plane networks goes back to Gilbert \cite{GILBERT} in the
early 60's. In comparison to  the well-known and 
well-studied random Bernoulli graphs\footnote{E.g. there are more than
800 references in \cite{Bollobas}.} $\G(n, p)$~
\cite{Bollobas,ER60,JLR2000}, only few papers
considered the probabilistic modeling of the communication
 graph properties of wireless sensor or ad hoc networks and the theory of 
\textit{random geometric graphs} (RGG)
 is still in development
with many problems left open (cf. \cite{PENROSE-RGG}).
 In contrary to the Bernoulli graphs, in
random  geometric graphs the probability of edge 
occurrences are not independent making them more difficult to study.
In RGG, a set of $n$ points are scattered, following a given
distribution,  in the deployment region
$\REGION$. Penrose studied the longest edge of 
the Euclidean  minimum spanning trees (MST) and 
 the longest nearest-neighbor link \cite{PENROSE2}. The same 
author established also that if the region of deployment
is a $d$-dimensional cube then the graph of communication
becomes $k$-connected as soon as its minimum degree reaches $k$
\cite{PENROSE}. In a slightly different setting,
Gupta  and Kumar \cite{Kumar-Gupta} stated that when
$n$ nodes are distributed uniformly in the disk of
unit area and their transmitting range is set to
$r = \sqrt{\frac{\ln{n} + c(n)}{\pi \, n}}$
 then the resulting network is connected with high probability
\footnote{An event $\xi_n$ is said to occur \textit{with high probability} or
\textit{asymptotically almost surely} (a.a.s. for short) if its probability
tends to $1$ as $n \ten \infty$. Formal definitions will be shortly given.}
if and only if $c(n) \ten  \infty$. Their results are obtained
making use of the theory of continuum percolation \cite{MEESTER-ROY},
 which is also used in \cite{DOUSSE} to investigate 
the connectivity of hybrid ad hoc networks.

As far as we know, the critical transmitting range and the critical coverage
range have been investigated first in \cite{PHILIPS-PANWAR-TANTAWI}
 for the case when the nodes are distributed in a square according
to a Poisson point process of fixed intensity. 
For the line of a given length, it has been studied in \cite{PIRET}.
% More stringent results 
%for both connectivity and coverage 
%for planar processes have been settled in \cite{Vlad}.
In \cite{SHAKKOTTAI}, Shakkottai \textit{et al.} considered 
 connectivity, coverage and hop-diameter of a particular sensor grid.

In this paper, we consider a model similar to those of
\cite{PHILIPS-PANWAR-TANTAWI,PIRET,Vlad}, but for three-dimensional
 settings. Furthermore, our results are illustrated with 
 two fundamental distributed protocols intended for the neighborhood
discovery and the code assignment problem \cite{BATTITI,COLORS}.
In particular, our results permit to do the average-case based analysis
 of the code assignment problem which has been investigated 
empirically in \cite[Section 4]{BATTITI}.

The remainder of this paper is organized as follows. Section 3 presents
 formal descriptions of the models used throughout the paper.
 Section 4 offers results about the relationships between the
sensing (resp. transmitting) range, the number of nodes and the
volume of the region to be monitored. In particular, we show 
how to quantify the minimum and maximum degrees of the network.
We also show how to compute  the hop-diameter of the underlying graph
 (defined as the maximum number of hops between any pair of sensor nodes),
when the transmitting range is slightly greater than the one required
 to have a connected network.
In Section 5, we present general algorithmic schemes for 
distributed protocols related to
our settings. In particular, we consider how to 
design a polylogarithmic protocol to allow the
nodes to discover their neighborhoods asymptotically almost surely.
Next, we turn on the design and analysis of a protocol
to assign codes in such random networks. The code assignment
problem consists to color the nodes of a graph in such a way
that any two adjacent nodes are assigned two different colors.

\section{Preliminaries and models}
To analyze the topology of sensor networks, three fundamental
models are needed: (a) a model for the spatial node distribution,
(b) a model for the wireless channel of communication between the nodes and (c)
 a model to represent the region monitored by a particular sensor.
Throughout this paper, nodes are randomly deployed in a
subset $\REGION$ of ${\mathbb{R}}^{3}$ following a uniform distribution.
To model the wireless transmission between the nodes, a radio
link model is assumed in which each node has a certain communication range,
denoted $\TRANS$. Two nodes are able
to communicate if they are within the transmitting range denoted $\TRANS$ of each other. 
Only bidirectional links are considered. 
Clearly, one imposes a graph structure
by declaring any two of the stations that lie within a 
given transmission range to be
 \textit{connected} by an \textit{edge}. The resulting
graph is called the \textit{reachability graph} and is denoted 
$\G(\rho,\, \TRANS)$ where $\rho = \frac{n}{\VOLUME}$ represents
the (expected) density of the nodes per unit volume. 
The {\it network coverage}
is defined as follows. For a node located at a point $p$,
 its monitored region is represented by a sphere, the \textit{sensing sphere}, 
of radius $\SENSE$ and centered at $p$. A region $\REGION$
is said \textit{covered} if every point in $\REGION$ is at distance
at most $\SENSE$ from at least a node.
We assume that every node has the same sensing range $\SENSE$ and the same transmitting range $\TRANS$.
Given the increasing interest in sensor networks, our purpose
is to design a solution where the nodes can maintain
both sensing coverage and  network connectivity.

\noindent
Throughout this paper, $\VOLUME$ always represents the volume of the region $\REGION$ to monitor
and we allow both $n$, $\VOLUME \ten \infty$,
in such a way that $n/\VOLUME \ten \rho$ (dense networks).
%The limiting process is
%well approximated by the homogeneous Poisson process of intensity $\rho \sim \frac{n}{\VOLUME}$.
Concretely, $\rho = \frac{n}{\VOLUME}$ represents the expected number of nodes per unit volume.

\medskip

\subsection*{Notations}
\begin{itemize}
\item {\bf Degrees of the reachability graph.} The degree of a node $v$ represents the number of
its neighbors in the graph $\G(\rho,\, \TRANS)$ 
and is denoted $d_v$. We consider here questions such as:
What is the required value of the transmitting range $\TRANS$
to have a reachability graph with a given  minimum (resp. maximum) degree $\delta$ (resp. $\Delta$)?  
\item {\bf Diameter.} The hop distance between two nodes $u$ and $v$ is defined
as the length of the shortest path (with respect to the number of hops) between them. 
The \textit{diameter} (or \textit{hop-diameter}) of a graph
is the maximum of minimum hop distances between any two pair of nodes.
Under the same hypothesis as above, what is the typical 
diameter of the reachability graph of the random sensor networks?
\item {\bf Degrees of coverage.} Define any convex region of ${\mathbb{R}}^{3}$ as having a
degree of coverage $k$ (i.e., being $k$-\textit{covered}) if every point of
the considered region is covered by at least $k$ nodes. Given the volume $\VOLUME$ of $\REGION$,
what is the required value of the sensing range $\SENSE$
to achieve a specified coverage degree $k$, $k>0$? 
\end{itemize}

 Throughout this paper, we will use 
standard mathematical notations \cite{De Bruijn}
 concerning the asymptotic
behavior of functions. We have
\begin{itemize}
\item $f(n)=O(g(n))$ if there exists a constant $c$ and a value $n_0$
such that $f(n) \leq c \, g(n)$ for any $n \geq n_0$.
\item $f(n)= \Theta(g(n))$ if there exists constants $c_1$ and $c_2$ and
a value $n_0$ such that $c_1 g(n) \leq f(n) \leq c_2 g(n)$ for any
$n \geq n_0$.
\item $f(n) = o(g(n))$ or $f(n) \ll g(n)$ if $\frac{f(n)}{g(n)} \ten 0$
as $n \ten \infty$.
\item $f(n) \sim g(n)$ if and only if $\lim_{n\ten \infty} \frac{f(n)}{g(n)} = 1$.
Equivalently, $f(n) \sim g(n)$ if and only if $f(n)=g(n)+o(g(n))$.
\item We recall that an event $\xi_n$ (depending on the value of $n$) is
said to occur \textit{asymptotically almost surely} (a.a.s.) 
if its probability tends to $1$ as $n \ten \infty$. We also 
say $\xi_n$ occurs 
\textit{with high probability} (w.h.p.) as $n \ten \infty$.
\end{itemize}

\section{Fundamental characteristics of random sensor networks}
To generate the network, the $n$ sensor nodes are distributed
in the fixed region of volume $\VOLUME$. If $\rho=\frac{n}{\VOLUME}$ tends to
some constant (i.e., $\rho = O(1)$), this process is well approximated by
a Poisson point process of finite intensity $\rho$ (see \cite{HALL}). 
Note that this assumption is well suited for both theoretical point of view and in practice 
since by a suitable rescaling \cite{MEESTER-ROY} %\cite[pp 30--31]{MEESTER-ROY}
all the properties obtained in this paper can be reformulated. For instance,
 any realization $\G(\rho, \, \TRANS)$ using a transmission range $\TRANS$ and with expected number
of nodes $\rho$ per unit volume coincides with another realization $\G({\rho}', \, {\TRANS}')$~with
a transmission range $\TRANS'$ and intensity $\rho'$ provided that $\rho'=(\TRANS/\TRANS')^{3} \rho$.
This model is also well suited for faulty nodes since if each node is 
independently faulty with some probability $p$, the properties obtained here
remain valid with $n$ replaced by $p \times n$. 

\subsection{Connectivity regime and minimum transmission range} %
As a warm-up, we begin with a natural question that often
arises concerning the minimum value of $\TRANS$ required to achieve connectivity \cite{Kumar-Gupta}.
We have the following theorem~:
\begin{theorem} \label{SIMPLE_CONNECTIVITY}
Let $\REGION$ be a bounded and connected set of ${\E}^3$ of volume $\VOLUME$.
Suppose that $n$ nodes are placed in $\REGION$ according to the uniform distribution
and assume that $\rho = \frac{n}{\VOLUME} = O(1)$.
 The network formed by adding edge between nodes
of distance at most 
\beq
\TRANS = \sqrt[3]{3 (\ln{n} + \omega(n))/ 4 \pi \rho}
\label{TRANS1}
\eeq
is connected if and only if $\lim_{n \ten \infty} \omega(n) = \infty$.
\end{theorem}

\begin{proof}
By assumption, the distribution of the nodes can be approximated  (see \cite[Section 1.7]{HALL}) by a 
Poisson point process of intensity $\rho = \frac{n}{\VOLUME}$ which has
the following property. The probability $p_{\ell}(i)$ that a randomly chosen
node $i$ has ${\ell}$ neighbors is given by
\beq
\label{PROBA_NEIGHBORS}
p_{\ell}(i) = \frac{\L(4/3 \rho \pi \TRANS^3 \R)^{\ell}}{{\ell}!} \exp{\L(- 4/3 \rho \pi \TRANS^3\R)} \, .
\eeq
Thus, if $\TRANS=\sqrt[3]{3 (\ln{n} + \omega(n))/ 4 \pi \rho}$,
 as $n \ten \infty$ no sensor node is isolated with probability
close to
\ben
\prod_{i=1}^{n} \L(1-p_{0}(i) \R) &=& \L( 1-e^{-4/3 \rho \pi \TRANS^3} \R)^n %
 = \L( 1 - \frac{e^{-\omega(n)}}{n} \R)^{n} %
 = \exp{\L(n \ln{\L(1-\frac{e^{-\omega(n)}}{n} \R)} \R)} \cr
        & = & \exp{\L(-e^{-\omega(n)} + O\L(\frac{e^{-2\omega(n)}}{n} \R) \R)} \, .
\label{NO_ISOLATED}
\een
Therefore, there is a.a.s. no isolated nodes if and only if $\omega(n)$ tends to
infinity with $n$.
To prove that the graph is also connected, we argue as in \cite{Kumar-Gupta}.
More precisely, we make use of results from continuum percolation \cite{MEESTER-ROY}.
In percolation theory, nodes are distributed with Poisson intensity $\rho$ and 
as in our model, two nodes are connected iff the distance between them is less than 
$r(\rho)$. %%%%ICI PERCOLATION
 Denote by $\GPOISSON$ the resulting infinite graph. Theorem~6.3 
of Meester and Roy \cite{MEESTER-ROY} states that almost
surely $\GPOISSON$ has at most one infinite-order component
for $\rho \geq 0$. Thus, almost surely the origin 
(the node distribution is conditioned on the origin having a node)
in $\GPOISSON$ lies in either an infinite-order component or
 is isolated. Our problem can be approximated by regarding
 that process as the restriction to $\REGION$ of the Poisson
process of intensity $\rho$ on ${\E}^{3}$. Denote by 
$\GPX$ this restriction of $\GPOISSON$ to $\REGION$. By the
above observation (see also \cite[Section 3]{Kumar-Gupta} or \cite{DOUSSE}),
the probability that $\GPX$ is disconnected is asymptotically
the same as the probability that it has at least one isolated node.
For large $n$, the difference between the $\GPX$ and
our model can be neglected. Thus, to ensure connectivity
 it suffices that  there is no isolated vertex.
By (\ref{NO_ISOLATED}), 
this is only achieved upon setting $\TRANS = \sqrt[3]{3 (\ln{n} + \omega(n))/ 4 \pi \rho }$,
with $\omega(n)  \ten \infty$.
%\ENDPROOF
\end{proof}

\noindent \textbf{Remark.}
It is important to note here that 
%in contrary to many existing results,
%, where the connectivity
% regime is given for extremely regular regions (sphere, disk, square, cube, etc.),
Theorem \ref{SIMPLE_CONNECTIVITY} concerns {\it all connected regions} $\REGION$ of ${\mathbb{R}}^3$
of bounded volume  ($\VOLUME = | {\REGION} | < \infty$).

\subsection{Coverage and minimum sensing range} To study coverage properties, we need
some results from integral geometry \cite{SANTALO,MILES}. The following
 lemma is due to Miles \cite{MILES2}.
\begin{lemma} \label{TH:MILES} %\textbf{Miles \cite{MILES2}}
Let $\REGION$ be a bounded set of ${\E}^3$ of volume $\VOLUME$. Let
$\mathcal{P}$ be a point process of ${\E}^3$ of intensity
$\rho$. Suppose that each point $x$ of the point process $\mathcal{P}$
monitors a sphere centered at $x$ and of radius $r$.
Define $N(p)$ as the number of these spheres containing
$p \in {\E}^{3}$. For any set ${\SREGION} \subset {\E}^{3}$, denote respectively by 
$\hdown({\SREGION}) = \inf_{ p \in {\SREGION}} N(p)$ 
and by $\hup({\SREGION}) = \sup_{ p \in {\SREGION}} N(p)$.
Then, the following holds
\beq
\lim_{\VOLUME \ten \infty} \, \sup_{r>0} \, \Big| \Pr\left[\hdown({\REGION}) \geq j \right] %
- \exp{\Big(-3  \rho \pi^2/32 \, j(j+1) \, \VOLUME \, \varphi(j+1, \, 4 \pi \rho r^3/3) \Big)} \Big| \ten 0\, ,%
 \quad (j \in \mathbb{N})
\label{EQ:DOWN}
\eeq
and
\ben
 \lim_{\VOLUME \ten \infty} \, \sup_{r>0} \, \Big| \Pr\left[\hup({\REGION}) \leq \ell \right]
 -  \exp{\Big( - f(\ell, \rho, \VOLUME)\left(1-\varphi(\ell-1, 4 \pi \rho r^3/3)  \right) \Big)} \Big| \ten 0\, , %
 \quad (\ell \in {\mathbb{N}}^{\star}) \, 
\label{EQ:UP}
\een
with
\beq
f(\ell,\rho, \VOLUME) = \rho \, \left( 4+ 3/8 (\pi^2+16)(\ell -1)+ 3\pi^2/32 (\ell-1)(\ell-2) \right) \, \VOLUME
\eeq
and
\beq
\varphi(x,y) = e^{-y} \left( \sum_{i=0}^{x} \frac{y^i}{i!} \right) \, .
\label{FUNCTION_P}
\eeq
\end{lemma}

\bigskip

\noindent
Lemma \ref{TH:MILES} tells us about the limiting probability distribution
of the number of spheres covering a point $p$ of $\VOLUME$. 
For a given sensing range $\SENSE$, to ensure the total coverage of the region $\REGION$
 (that is every point of $\REGION$ is within the sensing range of at least one sensor), 
 it suffices to have $\hdown({\REGION}) \geq 1$. The following result establishes the
relation between $\SENSE$, $\VOLUME$ and $n$~:
\begin{theorem}
\label{FIRST_COVERAGE}
Let $\REGION$ be a bounded set of ${\E}^3$ of volume $\VOLUME$.
Assume that $n$ sensors, each of sensing range $\SENSE$, are
 distributed uniformly and independently at random in $\REGION$.
Suppose that $\rho = \frac{n}{\VOLUME}$ is constant.
% Let $\mathcal{P}$ be a 
%point process over $\REGION$ with constant
%intensity $\rho = n/\VOLUME$. Let 
Let
\beq
\SENSE = \sqrt[3]{ 3 \, \frac{\left( \ln{n} + \ln{\ln{n}} + \omega(n)\right) \, \VOLUME }%
{4 \, \pi \, n } } \, .
\label{SENSE1}
\eeq
With probability tending to $1$ as $n \ten \infty$ (and $\VOLUME = O(n)$),
 every point of $\REGION$ is monitored by at least one sensor
 if and only if $\omega(n) \ten \infty$.
\end{theorem}

\begin{proof}
We have to find
the value of $\SENSE$ (in terms of $n$ and $\VOLUME$)
 such that as $n \ten \infty$ (or $\VOLUME \ten \infty$),
the probability  $\Pr\left[\hdown({\REGION}) > 0 \right] \ten 1$.
By formula (\ref{EQ:DOWN}) in lemma \ref{TH:MILES}, 
this probability is well approximated by 
$ \exp{\left(-3  \rho \pi^2/16 \,  \VOLUME \, \varphi(2, \, 4 \pi r^3/3) \right)} $
(we used $j=1$ in the formula). Therefore, it suffices to have
$n \, \varphi(2, \, 4 \pi r^3/3) \ll 1$.
By setting $\SENSE = \sqrt[3]{ 3 \, \frac{\ln{n} + \ln{\ln{n}} + \omega(n)}{4 \, \pi \, \rho }}$,
we obtain 
\ben
n \, \varphi(2, \, 4 \pi \rho r^3/3)  & =  & n \, \varphi(2, \ln{n}+\ln{\ln{n}} + \omega(n)) \cr
  & = & n \, \exp{\left(-{\ln{n}-\ln{\ln{n}} - \omega(n)}\right)} \cr
& \times & \left( 1+ {\ln{n}+\ln{\ln{n}} + \omega(n)} + %
\frac{1}{2} \left({\ln{n}+\ln{\ln{n}} + \omega(n)}\right)^2 \right) \cr
& = & O \left( \frac{\ln{n}^2 + \omega(n)^2} {\exp{(\omega(n))}} \right)  \, .
\een
Hence,  $\Pr\left[\hdown({\REGION}) > 0 \right] \ten 1$ if and only if $\omega(n) \ten \infty$.
\end{proof}

\noindent \textbf{Remark.}
Formulae (\ref{TRANS1}) and (\ref{SENSE1}) indicate that the radius $\SENSE$
required to achieve a sensing-covered network is greater than
the transmission range $\TRANS$ required to have a connected network.
For randomly deployed sensor nodes,
 these results show that if the transmission/sensing radius 
can be ``compared'', 
 connectedness arises slightly before sensing coverage. It is also important to note
that the arguments used to prove these Theorems show that both connectivity and
coverage  properties are subject to abrupt change known as phase transition phenomena. 
From an engineering standpoint, these formulae are crucial: For instance, (\ref{TRANS1}) indicates that
there is a critical transmission  effort required to each individual
node to ensure with high probability that any pair of nodes in the network can communicate with 
each other through multihop paths.
We refer here to the paper of
Goel \textit{et al.} \cite{STOC2004} and to Krishnamachari \textit{et al.} \cite{KRISHNA1}
for results about phase transition phenomena in wireless networks and random
geometric graphs. In particular, the authors of \cite{STOC2004} proved that
monotone properties for random geometric graphs have sharp thresholds 
(we refer to Friedgut and Kalai \cite{FRIEDGUT} for the classification of critical thresholds phenomena in
random graphs).

\subsection{Degrees and coverage}
In various networking problems, multiple-paths between any pair of 
two nodes are of importance. On one hand, the existence of 
multiple independent paths plays crucial role (e.g. in flooding
or in gossiping protocols).
On the other hand, in dense networks where each node has a 
great number of neighbors, the number of interferences 
makes the scheduling of communications difficult.

Similarly, different applications may require different degrees of sensing coverage.
While some protocols require that every location
in the considered region be monitored by one node, other
applications need significantly higher degrees of coverage.
In these directions, one is first interested in $1$-coverage
(which is solved by Theorem \ref{FIRST_COVERAGE})
but also in $k$-coverage for several values of $k$, in particular
for unbounded value of $k$, viz.  $k = k(n) \gg 1$.
Our aim in this paragraph is to address issues related to the probabilistic
relationships between the transmission/sensing ranges, the number of nodes
and the size of the region of deployment.  %In the following, we address these issues. 
More precisely, we are interested in values of the sensing range $\SENSE$ 
such that the event ``each point $p$ of the considered region $\REGION$
is covered by at least $k$ spheres of radius $\SENSE$'' occurs asymptotically 
almost surely. Similarly, we are interested in several degrees of connectedness
 depending on the transmission range. 

\begin{theorem}
\label{MULTIPLE_COVERAGE}
Let $\REGION$ be a bounded subset of ${\E}^3$ of volume $\VOLUME$.
Assume that $n$ sensors are deployed uniformly and independently at
random in $\REGION$ and $\rho = \frac{n}{\VOLUME} = O(1)$. 
Then as $n \ten \infty$, the following holds~: 
\begin{itemize}
\item[\textbf{(i)}] Given any fixed integer $\ell >0$, if the sensing range $\SENSE$ satisfies
$4/3 \, \pi \, \rho \, \SENSE^3 = \ln{n}+ \ell \ln\ln{n} + \omega(n)$ 
 then $\REGION$ is a.a.s. $\ell$-covered if and only if 
$1 \ll \omega(n) \ll \ln\ln{n}$. 
\item[\textbf{(ii)}] Given any function $c(n)$ satisfying $1 \ll c(n) \ll \frac{\ln{n}}{\ln\ln{n}}$
if $4/3 \, \pi \, \rho \, \SENSE^3 = \ln{n} + c(n) \ln\ln{n}$ then each point of
$\REGION$ is a.a.s. monitored by at least $\sim \, c(n)$ spheres and
at most $\sim \, e \, \ln{n}$ spheres.
\item[\textbf{(iii)}]  For any constant real number $\ell$, $\ell >0$,
 if  $4/3 \, \pi \, \rho \, \SENSE^3 = (1+\ell) \, \ln{n}$ then
  the number $N(p)$ of 
sensors covering and monitoring
each point $p$ of $\REGION$ satisfies a.a.s.
\beq
    - \, \frac{ \ell \,\ln{n}}{W_{-1}\L(- \frac{\ell}{e \, (1+\ell)}\R)} +
o\L( \ln{n} \R) \leq N(p) \leq  - \, \frac{ \ell \,\ln{n}}{W_{0}\L(- \frac{\ell}{e \, (1+\ell)}\R)} +
o\L( \ln{n} \R) \, ,
\label{W0W1}
\eeq
where $W_0(z)$ denotes the branch\footnote{See the Appendix for %
details on the Lambert W function \cite{LAMBERTW}.} of the Lambert W function $W(z)$
satisfying $-1 \leq W(z)$ and $W_{-1}(z)$ is the branch of the same function satisfying
$W(z) \leq -1$.
\item[\textbf{(iv)}] Given any function $c(n)$ satisfying $1 \ll c(n) \ll \frac{n}{\ln{n}}$,
if $4/3 \, \pi \, \rho \, \SENSE^3 = c(n) \ln{n}$ then each point $p$ of
$\REGION$ is a.a.s. covered by $N(p) \sim c(n) \ln{n}$.
\end{itemize}
\end{theorem}

\begin{proof} The proof of \textbf{(i)} is very similar to the one of
Theorem \ref{FIRST_COVERAGE} and is therefore omitted. The proofs
of \textbf{(ii)}, \textbf{(iii)} and \textbf{(iv)} 
rely on extremely precise analysis %of the behavior of the
of the truncated gamma function $p$ present in equations
(\ref{EQ:DOWN}) and (\ref{EQ:UP}) above. More specifically,
the gamma function is defined as
\[
\Gamma(x) = \int_{0}^{\infty} e^{-t} t^{x-1} dt \, .
\]
The incomplete gamma functions arise from the integral above by
decomposing it into two integrals, the first one from $0$ to $y$
and the second from $y$ to $\infty$~:
\ben
\gamma(x,y) & = & \int_{0}^{y}  e^{-t}\, t^{x-1} dt \, , %
\quad \quad \, {\EuFrak{{Re}}} (x) > 0 \, ,\cr
\Gamma(x,y) & = & \int_{y}^{\infty}  e^{-t}\, t^{x-1} dt \, , %
\quad \quad |\arg{y}| < \pi \, .%
\label{SPLIT-GAMMA}
\een
Specific cases are obtained when $x$ is an integer. For $x \in {\mathbb{N}}$,
we have
\beq
\Gamma(x+1,y) = x! \, e^{-y} \, \L(\sum_{k=0}^{x} \frac{y^k}{k!}\R) \, ,
\label{TRUNC-EXP}
\eeq
Thus, for any natural number $y$ 
the function $\varphi$ described in (\ref{FUNCTION_P}) can be expressed as
\beq
\varphi(x,y) = \frac{\Gamma(x+1,y)}{\Gamma(x+1)} \, .
\eeq
Uniform asymptotic expansions for the so-called incomplete
gamma function are now required to cope with the value of
$\varphi$ present in (\ref{EQ:DOWN}), resp. (\ref{EQ:UP}). These
expansions were derived by Temme \cite{TEMME} (see also 
\cite{WALTER} for a survey on incomplete gamma functions). 
By using the integral representation of
the truncated gamma, viz.
\beq
 \frac{\Gamma(x,y)}{\Gamma(x)} = \frac{e^{x \, \phi(\beta)}}{2 \pi i} %
\int_{c-i\infty}^{c+i\infty} e^{x \phi(t)} \frac{dt}{\beta-t} \, , \qquad %
0 < c < \beta \, ,
\label{START-TEMME}
\eeq
where
\beq
\phi(t) = t - 1 - \ln{t} \qquad \mbox{and} \qquad \beta = \frac{y}{x} \, ,
\label{SADDLE-POINT}
\eeq
the author of \cite{TEMME} arrives at an asymptotic representation 
of the form
\ben
\frac{\Gamma(x,y)}{\Gamma(y)} 
    & = & \frac{1}{2} \erfc{\L( \eta \, %
\sqrt{\frac{x}{2}} \R)} \cr
  & + & \frac{e^{-\frac{1}{2} x \eta^2}} {\sqrt{2 \pi x}} \,%
\L(\frac{1}{\beta - 1} - \frac{1}{\eta} \R) + %
\frac{e^{-\frac{1}{2} x \eta^2}} {\sqrt{2 \pi x}} \, 
 \L(\sum_{i=1}^{\infty}\frac{c_i(\eta)}{x^i}\R)\, .
\label{ASYMPTOTIC-EXP}
\een
In (\ref{ASYMPTOTIC-EXP}), the $\erfc$ is the complementary error
function defined by
\beq
\erfc(x) = 1-\frac{2}{\sqrt{\pi}} \int_{0}^{x} e^{-t^2} \, dt  \, ,
\label{ERFC}
\eeq
the coefficients $c_i(\eta)$ can be computed by a power series expansion
(we refer to \cite{TEMME} for details) and
\beq
\eta = (\beta -1) %
\sqrt{ 2 \, \frac{(\beta -1 - \ln {\beta)}}{(\beta-1)^2}} \, , \qquad \beta = \frac{y}{x} \, .
\label{ETA}
\eeq
In particular, as $x \ten \infty$ if $\eta$ is bounded the remainder terms of
(\ref{ASYMPTOTIC-EXP}) is exponentially small. Therefore, we have
\ben
\frac{\Gamma(x,y)}{\Gamma(y)} 
    & = & \frac{1}{2} \erfc{\L( \eta %
\sqrt{\frac{x}{2}} \R)} + 
\frac{e^{-\frac{1}{2} x \eta^2}} {\sqrt{2 \pi x}} %
\L(\frac{1}{\beta - 1} - \frac{1}{\eta} \R) + %
O\L( \frac{e^{-\frac{1}{2} x \eta^2}} {\sqrt{2 \pi x^3}}\R) \, .
\label{REGROUP}
\een
We quote here that computations of the functions $c_i$, involved in
(\ref{ASYMPTOTIC-EXP}), as well as the methods to get them
 are detailed in \cite{TEMME,TEMME2}.
Now using (\ref{ASYMPTOTIC-EXP}) and letting the 
value of $j$ (resp. $\ell$) vary,  we can make the values
\beq
\exp{\left(-3  \rho \pi^2/32 \, j(j+1) \, \VOLUME%
 \, \varphi(j+1, \, 4 \pi \rho r^3/3) \right)} 
\label{EXP1}
\eeq
 in (\ref{EQ:DOWN}) and its counterpart
\beq
%\exp{\left( - f(\ell, \rho, \VOLUME)\left(1-\varphi(\ell-1, 4 \pi \rho r^3/3)  \right) \right)}
exp{\left(- \rho \, \left( 4+ 3/8 (\pi^2+16)(\ell -1)+%
 3\pi^2/32 (\ell-1)(\ell-2) \right) \, %
\VOLUME\left(1-\varphi(\ell-1, 4 \pi \rho r^3/3)  \right) \right)}
\label{EXP2}
\eeq
 in (\ref{EQ:UP}) very close to $1$. In fact, suppose for instance that
$4/3 \, \pi \, \rho \, \SENSE^3 = \ln{n} + c(n) \ln\ln{n}$ as
given in the hypothesis of \textbf{(ii)}. Therefore,
$\varphi(j+1,4/3 \, \pi \, \rho \, \SENSE^3) = \varphi(j+1,\ln{n}+c(n)\ln\ln{n})$.
Since the expression inside the exponential in (\ref{EXP1}) must
tend to $0$, the value of 
$ \varphi(j+1,\ln{n}+c(n)\ln\ln{n})$ must behave like
\beq
 \varphi(j+1,\ln{n}+c(n)\ln\ln{n}) = \frac{1}{j^2 n \, \ln{n}^{\alpha}} , %
\qquad ( \alpha > 0) \, .
\label{MUST_BEHAVE}
\eeq
Using the fact that,
\ben
\erfc(x) = \frac{e^{-x^2}}{\sqrt{\pi} \, x}%
\L( 1+ O\L(\frac{1}{x^2}\R) \R), \quad x \ten \infty
\label{ERFC1}
\een
%and that $\jmin \ten \infty$, 
with the help of formula (\ref{REGROUP}),
 we then have to solve (asymptotically) with respect to $j$
\ben
\frac{1}{\jmin^2 \, n \, \ln{n}^{\alpha}  }
%& = & \frac{1}{2} \erfc{\L( \eta \, %
%\sqrt{\frac{(\jmin+1)}{2}} \R)} + 
%\frac{e^{-\frac{1}{2} (\jmin+1) \eta^2}} {\sqrt{2 \pi (\jmin+1)}} \,%
%\L(\frac{1}{\beta - 1} - \frac{1}{\eta} \R) + %
%O\L( \frac{e^{-\frac{1}{2} (\jmin+1) \eta^2}} 
%{\sqrt{2 \pi (\jmin+1)^3}}\R) \cr
&  = & %
\frac{e^{ -\frac{1}{2}(\jmin+1) \eta^2 }}{\sqrt{2 \pi (\jmin+1)}} %
\L( \frac{1}{\beta -1} \R) \, \L( 1 + O\L( \frac{1}{\jmin}\R) \R) \, ,
\label{GROS-BETA}
\een
which implies that $\beta > 1$.
%We also remark that the error term from (\ref{TO-SOLVE2}) can be neglected.
Equivalently, we have to find $\jmin$ satisfying
\ben
\sqrt{\frac{2 \pi}{\jmin^3}} \, \frac{(\beta - 1)}{n \, \ln{n}^{\alpha}}
\L( 1+O\L(\frac{1}{\jmin}\R) \R) = \exp{\L( -\frac{1}{2}(\jmin+1) \eta^2\R) } 
\, .
\een
Using (\ref{ETA}) and taking the logarithm 
of the previous equation, we obtain
\ben
(\jmin+1) \, (\beta -1 -\ln{\beta})  -   \frac{3}{2} \ln{\jmin} =
 \ln{n} + \alpha \, \ln{\ln{n}} 
- \ln{(\beta-1)} +O\L(1\R) \, .
\label{TO-SATISFY}
\een
Consequently, we then have~:
\ben
\jmin & = & \L(\frac{1}{\beta -1 - \ln{\beta}}\R)%
 \ln{n}+O\L( \frac{\ln{\ln{n}}}{\beta} \R)  + O\L(\frac{\ln{\beta}}{\beta}\R) + O\L(\frac{\ln{j}}{\beta} \R)  \, .
\label{TO-SOLVE3}
\een
%(We used $4/3\, \pi \, \rho \, \SENSE^3 = \ln{n} + c(n) \, \ln{\ln{n}}$
%with $1 \ll c(n) \ll \ln{n}/ \ln{\ln{n}}$.) 
Clearly, the equation (\ref{TO-SOLVE3}) 
above can be solved asymptotically (with respect to $\jmin$).
To this end, we need results concerning the asymptotic behavior of the
Lambert W function \cite{De Bruijn}. Whenever $z$ approaches $0^{-}$, $W_{-1}$ has a complete
asymptotic series expansion provided by the Lagrange inversion theorem (see \cite{LAMBERTW} and
\cite{De Bruijn})
which starts as
\ben
W_{-1}(z) = \ln{(-z)} - \ln{(-\ln{(-z)})} + %
O\L( \frac{\ln{(-\ln{(-z)})} }{\ln{(-z)}} \R) , %
\, (z \ten 0 \, , z < 0)\,.
\label{ASYMPT_W-1}
\een
Substituting $\beta$ by $\frac{\ln{n}+c(n)\ln\ln{n}}{j+1}$ and using standard analysis,
 it yields for $\jmin$ in (\ref{TO-SOLVE3})~:
\ben
\jmin = - \frac{\omega \L(n\R) \, \ln{\ln{n}}} %
 { W_{-1}\L(-\frac{\omega\L(n\R) \, \ln{\ln{n}}}{e \, \L( \ln{n} %
 +\omega\L(n\R) \, \ln{\ln{n}} \R)} \R) }\L( 1+o(1) \R)%
       =  \omega(n) + o\L(\omega(n)\R) \, ,
\label{SOLVED3}
\een
where we used the function $W_{-1}(z)$ and
the asymptotic formula (\ref{ASYMPT_W-1})
for $W_{-1}(z)$ as $z \ten 0^{-}$.

\medskip
\noindent
Now, we turn on the upper-bound -- given by \textbf{(ii)} -- of the number of spheres monitoring 
each point of $\REGION$. We argue as above. Fix a constant $\alpha > 0$,
 not necessarily the same as in (\ref{MUST_BEHAVE}).
As for formula (\ref{MUST_BEHAVE}), the expression inside the exponential in
(\ref{EXP2}) can be made sufficiently close to $0$ if we find $\ell$ such that
\beq
\ell^2 \, n \, (1-\varphi(\ell-1, 4/3 \, \pi \, \rho \, \SENSE^3)) %
 = \frac{1}{\ell^2 \, n \, \ln{n}^{\alpha} } \, ,
\eeq
which implies that
\ben
\varphi(\ell - 1, 4/3 \, \pi \, \rho \, \SENSE^3)  & = & %
\frac{\Gamma(\ell,  4/3 \, \pi \, \rho \, \SENSE^3)}{\Gamma(\ell)} = %
1 -  \frac{1}{\ell^2 \, n \, \ln{n}^{\alpha}}  \, .
\label{TO-SOLVE4}
\een
Instead of (\ref{ERFC1}), we now have the following asymptotic expansion~:
\ben
\erfc{\L(-x \R)} = 2 - \frac{e^{-x^2}}{\sqrt{\pi}x} %
+ O\L(\frac{e^{-x^2}}{x^3}\R) \, , \quad x \ten \infty  \, .
\label{ERFC2}
\een
Then,  by setting 
$4/3 \, \pi \, \rho \, \SENSE^3 = \ln{n} + c(n) \ln{\ln{n}}$, 
we find
\ben
\ell = e \, \ln{n} + o\L( \ln{n} \R) \, 
\label{SOLVED4}
\een
as stated for the upper-bound given by  the statement of
property \textbf{(ii)} in the Theorem.
The prove \textbf{(iii)}, we first remark that for any function $\omega(n) \gg 1$, the functions
$W_{0}$ and $W_{-1}$ involved above verify
\beq 
- \frac{  \omega(n) }{W_{-1}\L(- \frac{\omega(n)}{e \, (1+\omega(n))}\R)} \sim
- \frac{  \omega(n) }{W_{0}\L(- \frac{\omega(n)}{e \, (1+\omega(n))}\R)} \sim
\omega(n), \, \qquad n \ten \infty.
\eeq
Denoting by $j_1$ the at least number of spheres monitoring each point of
$\REGION$ when  setting the sensing range to $4/3 \, \pi \, \rho \, \SENSE^3 = (1+ \ell) \ln{n}$,
 instead of (\ref{TO-SOLVE3}) we have to solve
\ben
j_1 = \L(\frac{(1+\ell)\ln{n}}{j_1+1} - 1 + %
\ln{\L(\frac{(1+\ell)\ln{n}}{j_1+1} \R)} \R)^{-1} \, \ln{n} + O\L(\ln{\ln{n}}\R)%
 \, ,
\label{TO-SOLVE5}
\een
Equation (\ref{TO-SOLVE5}) can be solved asymptotically (w.r.t. $j_1$) and
this time we find
\ben
j_1 =  - \, \frac{ \ell \,\ln{n}}{W_{-1}\L(- \frac{\ell}{e \, (1+\ell)}\R)} %
+ o\L( \ln{n} \R) \, .
\label{J1J1}
\een
Similarly, denote by $j_2$ the at most number of spheres covering each point of
$\REGION$, we can argue as (\ref{SOLVED4}) to find $j_2$~:
\ben
j_2 =  - \, \frac{ \ell \,\ln{n}}{W_{0}\L(- \frac{\ell}{e \, (1+\ell)}\R)} +
o\L( \ln{n} \R) \, .
\label{J2J2}
\een
%\ENDPROOF
Combining (\ref{J1J1}) and (\ref{J2J2}),
we just have the property \textbf{(iii)} as stated.
Similarly, one can prove the property \textbf{(iv)} of the Theorem.
\end{proof}

Observe that the previous results reflect also the degrees
of the nodes of the reachability graph simply by replacing the sensing range $\SENSE$
with the transmitting range $\TRANS$ and we have the following~:
\begin{corollary}\label{MULTIPLE_DEGREE}
Let $\REGION$ be a subset of ${\E}^{3}$ such that $|{\REGION}| = \VOLUME < \infty$
and assume that $n$ sensor nodes are deployed in $\REGION$ following a uniform distribution.
Suppose that $\frac{n}{\VOLUME}$ is a constant $\rho$.
Denote by $\delta$ (resp. $\Delta$) the minimum (resp. maximum) degree of a given network
and for any node $v$, $d_v$ denotes the degree of $v$.
The network formed by adding edge between nodes of distance at most $\TRANS$ have the
following properties asymptotically almost surely as $n \ten \infty$~:
\begin{itemize}
\item[\textbf{(i)}] For any fixed integer $\ell > 0$, if the transmitting range 
$\TRANS$ satisfies $4/3 \, \pi \, \rho \, \TRANS^3 = \ln{n}+ \ell \ln\ln{n} + \omega(n)$ 
 then the minimum (resp. maximum) degree $\delta$ (resp. $\Delta$) 
of the reachability graph satisfies $\delta \geq \ell+1$ (resp. $\Delta \leq e \ln{n}$).
\item[\textbf{(ii)}] For any function $c(n)$ s.~t.  $1 \ll c(n) \ll \frac{\ln{n}}{\ln\ln{n}}$
if $4/3 \, \pi \, \rho \, \TRANS^3 = \ln{n} + c(n) \ln\ln{n}$ then each node $v$
of the network has a degree $d_v$ comprised between $c(n)$ and
 $e \, \ln{n}$.
\item[\textbf{(iii)}]  For any constant real number $\ell$  ($\ell >0$),
 if  $4/3 \, \pi \, \rho \, \TRANS^3 = (1+\ell) \, \ln{n}$ then the degree $d_v$ of any node 
$v$ of the network satisfies
\beq
    - \, \frac{ \ell \,\ln{n}}{W_{-1}\L(- \frac{\ell}{e \, (1+\ell)}\R)} +
o\L( \ln{n} \R) \leq d_v \leq  - \, \frac{ \ell \,\ln{n}}{W_{0}\L(- \frac{\ell}{e \, (1+\ell)}\R)} +
o\L( \ln{n} \R) \, ,
\label{W0W1-bis}
\eeq
\item[\textbf{(iv)}] Given any function $c(n)$ s.~t. $1 \ll c(n) \ll \frac{n}{\ln{n}}$,
if $4/3 \, \pi \, \rho \, \TRANS^3 = c(n) \ln{n}$ then each node $v$ of
the network has a degree $d_v \sim c(n) \ln{n}$.
\end{itemize}
\end{corollary}

\begin{proof}
For each one of the statements \textbf{(i)}, \textbf{(ii)}, \textbf{(iii)} and
\textbf{(iv)}, by substituting $\SENSE$ in Theorem  \ref{MULTIPLE_COVERAGE} with 
$\TRANS$,  each geographical point $p$ of $\REGION$ 
lies inside the corresponding number of ``spheres of communications''.
 In particular, the points representing the centers of 
the sensors are asymptotically inside  the same number of spheres. Therefore,
 the proof of the statements \textbf{(i)}, \textbf{(ii)}, \textbf{(iii)} and
\textbf{(iv)} in the Corollary follows the previous proof of Theorem 
 \ref{MULTIPLE_COVERAGE}.
\end{proof}

\subsection{Hop-diameter}
In this paragraph, we consider a setting slightly different from the above.
For sake of simplicity, we consider a \textit{cubic} region $\REGION$ of 
the Euclidean space ${\E}^{3}$. In what follows, the transmission 
range  of the nodes is set to a certain value
$\TRANS$ such that 
\ben
\lambda = \lim_{n\ten \infty} \frac{ 4/3 \, \pi \, n \, \TRANS}{ \VOLUME \, \ln{n} }%
 \,, \quad \lambda \in (0, \, \infty] \,.
\label{DEF-REGIME}
\een
%\noindent \textbf{Notations.} 
For $\lambda$ defined 
by (\ref{DEF-REGIME}) and borrowing terms from 
Bernoulli random graphs $\G(n,p)$ \cite{Bollobas,ER60,JLR2000}, 
with respect to the connectivity property, the regime is referred to as~:\\
\indent $\bullet$  the {\it subcritical} regime  if $\lambda<1$, \\
\indent $\bullet$  the {\it critical} regime if $\lambda=1$ and \\
\indent $\bullet$  the {\it supercritical} regime if $\lambda>1$.\\
%Moreover, if the limit $\ell$ defined in (\ref{DEF-REGIME})
% equals $\infty$, the regime is said \textit{highly supercritical}.
%Assume that $n$ sensors are scattered in
%$\REGION$ following the uniform and independent distribution and suppose
For transmission ranges  of the form 
$\TRANS = \sqrt[3]{3/4 \pi \, (1+\ell) \ln{n}/n}$,
 that is the connectivity regime is supercritical, we have the following result
 for the order of the diameter of the network: 
\begin{thm} \label{DIAMETER}
Suppose that $n$ sensor nodes are randomly deployed in
 a cubic region of volume $\VOLUME$ of ${\E}^{3}$ according to the uniform
distribution. If their common transmission range is set to
$\TRANS = \sqrt[3]{\,\frac{3\,(1+\ell) \,\ln{n} \,\VOLUME}{4 \, \pi \, n}\, }$ 
with $\ell > 11/5$,
then the diameter $\DIAM$ of the network
satisfies~: 
\beq
\lim_{n \ten \infty} %
\Pr\coeff{ \DIAM \leq \, 12  \, \sqrt[3]{ \frac{\pi \, n}{6 \, (1+\ell) \ln{n}}} \, \, } %
= 1 \, .
\label{EQ:DIAMETER}
\eeq
\end{thm}

\begin{proof}
\vspace{-0.2cm}
\begin{center}
  \begin{figure}[h]
    \hfill
    \begin{minipage}[t]{3.5cm}
      \begin{center}
	\begin{pspicture}(0,0)(5,3)
%	  \psgrid
	  \psline[linewidth=1pt]{-}(0.8,2.1)(1.8,1)
	  \psline[linewidth=1pt]{->}(1.8,1)(4.1,1)
%	  \rput(1.5,1.5){$\frac{\sqrt[3]{\VOLUME}}{k}$}
	  \psfig{file=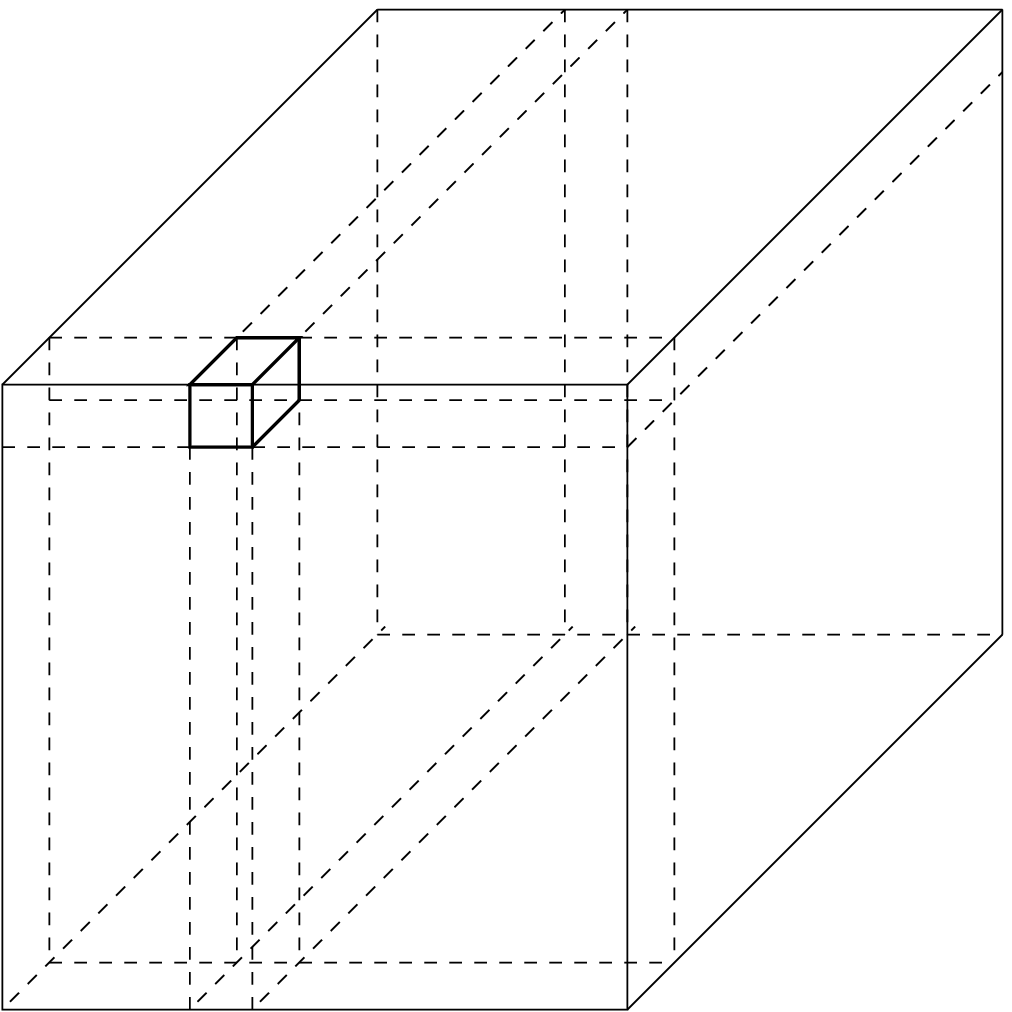,width=3.5cm,height=3.5cm,angle=0}
	\end{pspicture}
      \end{center}
      \label{FIG:DIAM}
    \end{minipage}
    \hfill
    \vspace{-0.2cm}
    \begin{minipage}[t]{3.3cm}
      \begin{center}
	\begin{pspicture}(0,0)(3,3)
%	  \psgrid
	  \psline[linewidth=2pt]{<->}(0,1.1)(2,1.1)
	  \rput(1,1.5){$\frac{\sqrt[3]{\VOLUME}}{k}$}
	  \psfig{file=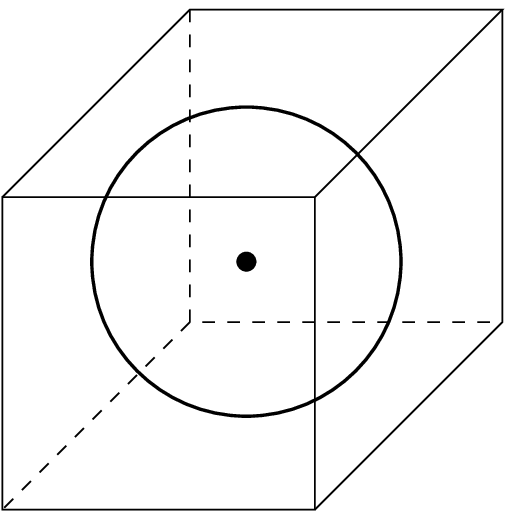,width=3.3cm,height=3.3cm,angle=0}
	\end{pspicture}
      \end{center}
      \label{FIG:SUBCUBE}
    \end{minipage}
    \hfill
    \begin{minipage}[t]{7.0cm}
      Split the cube of volume $\VOLUME$ into $k^3$ sub-cubes, 
${\REGION}_{1}$, $\cdots$, ${\REGION}_{k^3}$ of equal volume. 
Each of these sub-cubes has side $\sqrt[3]{\VOLUME}/k$. Choose
$k$ such that a sphere of radius $\TRANS$ can entirely fit inside a
sub-cube ${\REGION}_{i}$ (cf. figures).
    \end{minipage}
    \hfill
  \end{figure}
\end{center}
That is (under the hypothesis of the Theorem),
 $k = \frac{\sqrt[3]{\VOLUME}}{2 \TRANS} = %
\sqrt[3]{\frac{\pi \, n}{6 \, (1+\ell) \, \ln{n}} }$. For sake of simplicity,
 let us suppose that this value of $k$ is an integer. By (\ref{W0W1-bis})
of Corollary \ref{MULTIPLE_COVERAGE}, with high probability there is 
$\Theta(\ln{n})$ nodes inside the sphere of radius ${\VOLUME}^{1/3}/2 k$.
Consider now two adjacent sub-cubes as depicted in the figure
 below~:
\vspace{-0.2cm}
\begin{center}
  \begin{figure}[h]
    \hfill
    \begin{minipage}[t]{7.0cm}
      \begin{center}
	\begin{pspicture}(0,0)(8,4)
%	  \psgrid
	  \psline[linewidth=2pt]{->}(4.6,2.9)(3.3,1.5)
	  \rput(4.8,3.1){$L_1$}	  
	  \psfig{file=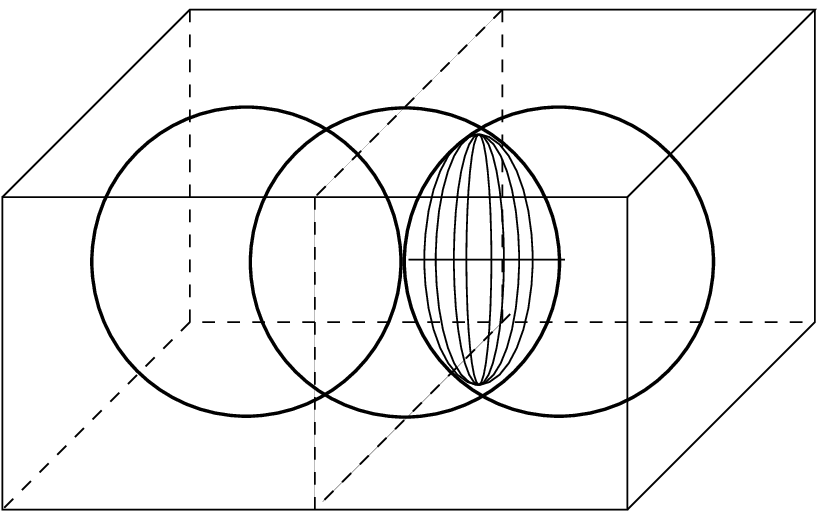,height=3.5cm,angle=0}
	\end{pspicture}
      \end{center}
      \label{FIG:INTERSECTION}
    \end{minipage}
    \hfill
    \vspace{-0.5cm}
    \begin{minipage}[t]{7.0cm}
The lens-shaped region, here denoted $L_1$, represents the intersection of two
spheres of radius $\TRANS$ and whose centers are at distance
$\TRANS$. A bit of trigonometry shows that the volume of
such intersection is given by $\frac{5 \, \pi \, {\TRANS}^{3}}{12}$.
    \end{minipage}
    \hfill
  \end{figure}
\end{center}
According to the uniform distribution, there is no node inside
each lens of volume $|L_1| = \frac{5 \, \pi \, {\TRANS}^{3}}{12}$
with probability
\beq
\left( 1 - \frac{|L_1|}{\VOLUME} \right)^n = %
\left( 1 - \frac{5 \, (1+\ell) \, \ln{n}}{16 \, n} \right)^n \, .
\eeq
Since each sub-cube has at most $6$ lenses similar to $L_1$,
none of these lenses is empty with probability at least
\ben
\left(1 - \left(1-\frac{5\,(1+\ell)\,\ln{n}}{16\, n}\right)^n\right)^{6 k^3}%
   &=& %
\left(1-\exp{\left(n\ln{\left(1-\frac{5\,(1+\ell)\,\ln{n}}{16\,n}\right)}\right)}\right)^{6 k^3} \cr
   & \geq & \left( 1 - \exp{\left(-\frac{5\,(1+\ell) \, \ln{n}}{16} \right)} \right)^{\frac{\pi \, n}{(1+\ell) \, \ln{n}}} \cr
   & \geq & \exp{\left({\frac{\pi \, n}{ (1+\ell) \, \ln{n}}} %
 \ln{\left(1 - \exp{\left(-\frac{5\,(1+\ell) \, \ln{n}}{16} \right)} \right)}\right)} \cr
   & \geq & \exp{\left(-\frac{\pi}{(1+\ell) \, \ln{n}}  \, %
\times \, \frac{1}{n^{\left(\frac{5 \, (1+\ell)}{16} - 1\right)}} \right)} \, .
\een
Therefore, if $\ell > 11/5$ each lens similar to $L_1$ is 
non-empty a.a.s. Hence, from two adjacent sub-cubes
${\REGION}_{i}$ and ${\REGION}_{j}$, communications
between any node $u \in {\REGION}_{i}$ and any
node $v \in {\REGION}_{j}$ need at most $6$ hops as shown
by the figure below
\vspace{-0.2cm}
\begin{center}
  \begin{figure}[h]
    \hfill
    \vspace{-0.5cm}
    \begin{minipage}[t]{7.0cm}
      \begin{center}
	\begin{pspicture}(0,0)(8,4)
%	  \psgrid
%	  \psline[linewidth=2pt]{->}(4.6,2.9)(3.3,1.5)
%	  \rput(4.8,3.1){$L_1$}	  
	  \psfig{file=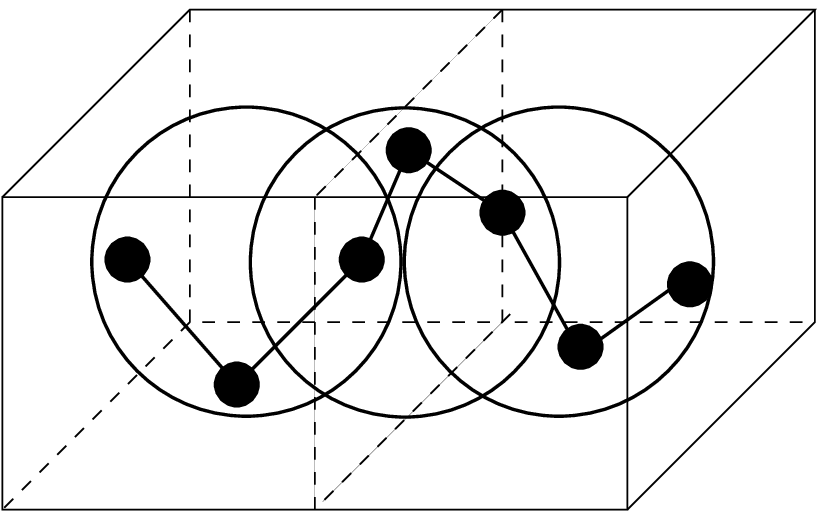,height=3.5cm,angle=0}
	\end{pspicture}
      \end{center}
      \label{FIG:INTERSECTION_END}
    \end{minipage}
    \hfill
    \begin{minipage}[t]{7.0cm}
      \bigskip
      \bigskip
      \bigskip
      \noindent
      By simple counting arguments, the proof of the Theorem is now complete.
    \end{minipage}
  \end{figure}
\end{center}
\end{proof}

\noindent \textbf{Remark.}  %
We conjecture that in the supercritical regime for connectivity,
that is for transmission ranges of the form 
$\TRANS = \sqrt[3]{3/4 \pi \, (1+\ell) \ln{n}/n}$ with $0 < \ell \leq 11/5$
the diameter of the network is (in probability) of order of magnitude 
$\Theta(n^{1/3}/ \ln{n}^{1/3})$. In contrary, it would be 
much more difficult to capture the diameter in the just critical
regime, viz. if $\TRANS = \sqrt[3]{3/4 \pi \, (\ln{n}+\omega(n))/n}$
with $\omega(n) \gg 1$.

\section{Distributed protocols}
In this section, we consider some distributed protocols
 which are built on the top of the previous results concerning
the main characteristics of random sensor networks.
As emphasized by the fundamental papers \cite{EMULATION,EXP-GAP,BASAGNI},
to cite only a few, the number of nodes $n$, the diameter $\DIAM$
and the maximum degree $\Delta$ of the networks can play 
 crucial role when designing distributed protocols for
radio networks. In fact, the executions of these
algorithms are often measured in terms of $n$, $\DIAM$ and $\Delta$.

\subsection{A model for protocols}
A commonly encountered model for distributed protocols
 is described briefly in this paragraph. We refer to \cite{EMULATION,BASAGNI}
for detailed descriptions of this model. 
A distributed protocol for multihop networks
is a protocol executed at each node in the network in the following
way~:
\begin{itemize}
\item The time of execution is considered to be slotted and
are subdivided into time slots or rounds.
\item In each round, a node acts either as transmitter or
as receiver. A node $u$ receives a message $m$ sent by one of its
neighbors in a given round if and only if (a) it acts as a receiver and
(b) exactly one of its neighbors acts as a transmitter. 
If two or more neighbors of $u$ are sending at the same time-slot
the node $u$ does not receive nothing. That is, nodes are unable
to distinguish between \textit{collision} and the lack of message.
\item The nodes are assumed to be distinguishable, that is each node
has an unique identifier, ID for short, ranging from $1$ to $n$.
In our settings, we assume also that the nodes are aware of 
their number $n$ as well as the volume $\VOLUME$ of the region
of interest.
\end{itemize}

\noindent \textbf{Remark.} In what follows and 
without loss of generality, we assume 
that the nodes can transmit 
messages up to a distance of order 
$O( \sqrt[3]{ \frac{\ln{n} \, \VOLUME}{n}} )$.
We note that if the (common) transmission range of the nodes
 is such that ${\TRANS}^{3} \gg \frac{\ln{n} \, \VOLUME}{n}$, 
 the results presented in the following paragraphs can be
easily extended using  the same global ideas.

\subsection{A simple protocol for neighborhood discovery}
The first distributed protocol which will be discussed is a protocol
called \textsc{ExchangeID}. This algorithm allows each node $u$
to discover the set of its neighbors, denoted $\Gamma(u)$. This protocol
 appears to be useful since as already stressed the nodes are deployed
in a random fashion and therefore, they do not have any \textit{a priori}
knowledge of their respective neighbors. The neighborhood discovery
is done using a simple randomized greedy algorithm 
 whose pseudo-code
 is given in the following~: \\
\noindent (0) \textbf{Protocol} \textsc{ExchangeID}($n$, $\VOLUME$, $\TRANS$)\\
(1) \indent \textbf{Begin} \\
(2) \indent \indent  Compute $\ell$ verifying :
$
\TRANS \times%
 \sqrt[3]{\frac{4 \, \pi \, n}{3 \, \ln{n} \, \VOLUME}} = 1 + \ell 
$ ; \\
(3) \indent \indent Then set %
$\Delta :=- \, \ell \times \ln{n}/ W_{0}\L(- \frac{\ell}{e \, (1+\ell)}\R)$~
 and $ C_{\ell} := 2 \, \exp{\left(2 \, W_{0}\L(- \frac{\ell}{e \, (1+\ell)}\R)\right)}$ ;\\
(4) \indent \indent \textbf{For} $i:=1$ to $\lceil C_{\ell} \ln{n}^2 \rceil $ \textbf{Do} \\
(5) \indent \indent \indent \textbf{With probability} $\frac{1}{\ln{n}}$, %
 each node $u$ sends a message containing its own ID ; \\
(6) \indent \indent \textbf{EndFor} \\
(7) \indent \textbf{End.} \\

\bigskip

\begin{thm} \label{TH_EXCHANGE}
Suppose that $n$ sensor nodes are randomly deployed in 
a region of volume $\VOLUME$ following a uniform distribution.
If their transmission range satisfies
$\TRANS \geq \sqrt[3]{\frac{3}{4 \, \pi \, n} \, %
\left[\ln{(n)}+\omega(n)\right] \VOLUME}$,
with $\omega(n) \gg 1$ but
$\TRANS=O\left(\sqrt[3]{\frac{3 \ln{(n)} \, \VOLUME}{4 \, \pi \, n}}\right)$,
 then after one execution of
 \textsc{ExchangeID}($n$, $\VOLUME$, $\TRANS$), with
probability tending to $1$ as $n \ten \infty$,  every node
has received correctly all the identities of all its neighbors.
\end{thm}

\begin{proof}
The proof of Theorem \ref{TH_EXCHANGE} relies on
two facts, viz., {\bf (1)} the main characteristics of the 
random Euclidean network and 
{\bf (2)} the number of iterations  $T= C_{\ell} \ln{(n)}^2$ in the 
main loop  of \textsc{ExchangeID} is 
sufficient for the nodes to send its ID {\it at least once} 
to all its neighbors. For the first point {\bf (1)}, we have seen that
for any node $v$ of the network, the degree of $v$ ($d_v = |\Gamma(v)|$)
satisfies w.h.p. (cf. Corollary \ref{MULTIPLE_DEGREE} equation 
(\ref{W0W1-bis}))~:
\beq
d_v \leq  - \, \frac{ \ell \,\ln{n}}{W_{0}\L(- \frac{\ell}{e \, (1+\ell)}\R)} +
o\L( \ln{n} \R) \, .
\eeq
Therefore, at the regime considered
in the hypothesis of Theorem \ref{TH_EXCHANGE}, 
 the maximum degree of the graph is
(with high probability) bounded by $c_{\ell} \ln{n}$ (where
$c_{\ell}$ satisfies e.g. $c_{\ell} = 2 \, {W_{0}\L(- {\ell}/{e \, (1+\ell)}\R)}$.
Using this latter remark, let us complete the proof of our Theorem.
For any distinct pair $(i,j)$ of adjacent nodes and
any time slot $t \in \left[1, \,  C_{\ell} \, \ln{(n)}^2 \right]$, 
define the random variable (r.v., for short) $X_{i \rightarrow j}^{(t)}$ as follows:
\beq
X_{i \rightarrow j}^{(t)} = 
\begin{array}\{{rrl}.
  \bullet &   1 &  \mbox{\textbf{if and only if} the node } j 
%          &     & 
 \mbox{ does not receive the ID of } i \\
          &     & \mbox{at time } t \in \left[1, \,  C_{\ell} \, \ln{(n)}^2 \right] \, , \\
  \bullet &   0 &  \mbox{\textbf{otherwise}} \, .
\end{array}
\label{XIJt}
\eeq 
In other terms, the set 
\beq
\left\{ X_{i \rightarrow j}^{(t)}, i,\, j \neq i,\, t \in \left[1, \, C_{\ell} \,\ln{(n)}^2  \right] \right\}
\eeq
 denotes a set of random variables 
 that counts the number of ``arcs'' $i \rightarrow j$ such that 
$j$ has never received the ID of $i$. Denote by $X$
the r.v. 
\beq
X = \sum_{i \neq j} X_{i \rightarrow j} \, 
\eeq
where $X_{i \rightarrow j}=1$ \textit{iff} $X_{i \rightarrow j}^{(t)} = 1$
for all $t \in  \left[1, \,  C_{\ell} \, \ln{(n)}^2 \right]$.
Now, we have the probability that $i$ does not succeed to send its ID to 
$j$ at time $t$:
%\ben
%& & \Pr{\left[  X_{i \rightarrow j}^{(t)} = 1  \right]}  =  \left( 1- \frac{1}{\ln{(n)}}\right) \cr
%& + &   \frac{1}{\ln{n}} \times \left( 1 - \left( 1-\frac{1}{\ln{n}} \right)^{d_j} \right)
%\een
\ben
 \Pr{\left[  X_{i \rightarrow j}^{(t)} = 1  \right]}  =  \left( 1- \frac{1}{\ln{(n)}}\right) % \cr
 +    \frac{1}{\ln{n}} \times \left( 1 - \left( 1-\frac{1}{\ln{n}} \right)^{d_j} \right) \, .
\een
%where $d_j(t)$ denotes the degree of $j$ at time $t$. 
Therefore, considering the whole 
range $\left[1, \, C_{\ell} \, \ln{(n)}^2 \right] $, after a bit of algebra we obtain
\ben
 \Pr{\left[  X_{i \rightarrow j} = 1  \right]} & \leq &  %
\left(1 - \frac{ e^{-c_{\ell}}}{\ln{(n)}}\right)^{\ln{(n)}^2 C_{\ell}} \cr
& \leq & \exp{\left( -\ln{(n)} e^{-c_{\ell}} C_{\ell} \right) }
\een
which bounds the probability that $i$ has never sent its ID to $j$
for all $t \in \left[1, \, \ln{(n)}^2 C_{\ell} \right]$.
By linearity of expectations and since by (\ref{W0W1-bis}) the number of
edges is of order $O(n \ln{n})$, we then have
\ben
\Esp{\left[X \right]} %& = & \sum_{i \neq j}  %
%\Esp{\left[X_{i \rightarrow j}\right]} \cr
  \leq  O\left( n \ln{(n)} \,%
 \exp{\left( -\ln{(n)} \, e^{-c_{\ell}} \, C_{\ell} \right) }\right) \, .
\een
Thus, $\Esp{\left[X \right]} \ll 1$ as $n \rightarrow \infty$ for a
certain constant $C_{\ell}$ such that, say 
\beq
C_{\ell} \geq 2 e^{c_{\ell}} \, .
\eeq
Note that this constant can be computed for any given $\ell$
using, e.g. $c_{\ell} = 2 \, {W_{0}\L(- {\ell}/{e \, (1+\ell)}\R)}$
 and using the first moment method \cite{ALON}, 
one completes the proof of Theorem \ref{TH_EXCHANGE}.
\end{proof}

\subsection{Coloring and codes assignment}
According to the rules of distributed protocols given above, when
a node is transmitting, all the nodes within its transmitting range
must be silent. Moreover in multihop environments, collisions can
also occur when two non-adjacent nodes are trying to transmit to
a common neighbor (\textit{hidden collisions}). To circumvent
 these problems, researchers use to assign orthogonal codes to
 the nodes, a problem equivalent to that of coloring the graphs 
associated to the physical network \cite{BATTITI,IRENE,COLORS}.

To assign codes to the nodes of the network, let us consider the following
simple and intuitive randomized protocol called \textsc{AssignCode}.
 Each vertex $u$ has an initial list of colors 
(also referred as \textit{palette})
of size $d_u +1 = |\Gamma(u)|+1$ and starts uncolored. We can
assume that each node knows its neighbors (or at least a great
part of them) by using the previous algorithm, viz. \textsc{ExchangeID}.
 Then, the
protocol \textsc{AssignCode} proceeds in rounds. In each round,
each \textit{uncolored vertex} $u$, simultaneously and independently 
picks a color, say, $c$ from its list. Next, the station $u$ attempts to send
this information to his neighborhood denoted by
$\Gamma(u)$. Trivially, this attempt succeeds iff 
there is no collision. Before attributing
the color $c$ definitely to $u$, its neighbors has to sent
one by one a message of reception. Note that this can be done
deterministically in time $O(\ln{n})$ 
since $u$ can attribute  to its active neighbors in $\Gamma(u)$ a 
 predefined ranking
ranging from $1$ to $|\Gamma(u)|$.  Therefore,
$u$ sends a message of \textit{confirmation} and its neighbors undergo
 an \textit{update} of their proper palettes and of their active neighbors.
Hence, at the end of such a round the new colored vertex $u$ 
 can quit the protocol. The details of 
\textsc{AssignCode} follows~:

\bigskip

\noindent( 0) \textbf{Protocol} \textsc{AssignCode}($n$, $\VOLUME$, $\TRANS$) \\
( 1) \indent \textbf{Begin} \\
( 2) \indent  \indent Each vertex $u$ has an initial \textit{palette} of colors, say
  $p(u) = \{c_1, \, c_2 , \, \cdots, c_{d_u+1} \}$; \\
( 3) \indent \indent  Compute $\ell$ verifying :
$
\TRANS \times%
 \sqrt[3]{\frac{4 \, \pi \, n}{3 \, \ln{n} \, \VOLUME}} = 1 + \ell 
$ ; \\
( 4) \indent \indent Then set %
$\Delta :=- \, \ell \times \ln{n}/ W_{0}\L(- \frac{\ell}{e \, (1+\ell)}\R)$~
 and choose e.~g.~ $ D_{\ell}:= %
-\frac{24 \, \ell}{ W_{0}\left(- \frac{\ell}{e \, (1+\ell) } \right)}$ ;\\
( 5) \indent \textbf{For} $i := 1$ to $D_{\ell} \, \ln{(n)}^2$ \textbf{Do} \\
( 6) \indent \indent  For each vertex $u$ do \\
( 7) \indent \indent \indent  $\bullet$ Pick a color $c$ from $p(u)$ ; \\
( 8) \indent \indent \indent  $\bullet$ Send a message containing $c$ %
 {\bf with probability} %
    $\frac{1}{\Delta + |p(u)|}$ ; \\
( 9) \indent \indent \textbf{If} {no collision} \textbf{Then} \\
(10) \indent \indent \indent%
  Every station $v$ in $\Gamma(u)$ gets the message properly ; \\
(11) \indent \indent \indent%
      One by one (in order) every member of $\Gamma(u)$ sends a message ; \\
(12) \indent \indent \indent%
      ($\star$ This step can be synchronized by always allowing 
           $\Delta$ time slots. $\star$) \\
(13) \indent \indent \textbf{EndIf} \\
(14) \indent \indent \textbf{If} {$u$ receives all the $|\Gamma(u)|$ messages} %
\textbf{Then} \\
(15) \indent \indent \indent%
      $u$ sends a message of \textbf{confirmation} and goes to sleep ; \\
(16) \indent \indent \indent%
      every station in $\Gamma(u)$ removes the color $c$ from its palette ; \\
(17) \indent \indent \textbf{EndIf} \\
(18) \indent \textbf{EndFor} \\
(19) \indent \textbf{End.}

\bigskip
\begin{thm} \label{TH_ASSIGN}
Suppose that $n$ sensor nodes are randomly deployed in 
a region of volume $\VOLUME$ following a uniform distribution.
If their transmission range satisfies
$\TRANS \geq \sqrt[3]{\frac{3}{4 \, \pi \, n} \, %
\left[\ln{(n)}+\omega(n)\right] \VOLUME}$,
with $\omega(n) \gg 1$ but
$\TRANS=O\left(\sqrt[3]{\frac{3 \ln{(n)} \, \VOLUME}{4 \, \pi \, n}}\right)$,
 then after one execution of
 \textsc{AssignCode}($n$, $\VOLUME$, $\TRANS$), with
probability tending to $1$ as $n \ten \infty$,  every pair of nodes
at distance at most $\TRANS$ from each other have received
two distinct codes (colors).
\end{thm}

\begin{proof}
 Although more complicated, 
the proof of Theorem \ref{TH_ASSIGN} is very similar to the
one of Theorem \ref{TH_EXCHANGE}.
For any distinct node $u$, recall that 
$\Gamma(u)$ represents the set of its
neighbors and denote by $p_u$ the size of its current palette. Now, define the
random variable $Y_{u}$  as follows:
\beq
Y_{u} = 
\begin{array}\{{rrl}.
  \bullet &   1 &  \mbox{\textbf{if and only if} the node } u, \\ 
  &   &  \mbox{ remains uncolored after the } D_{\ell} \ln{n}^2 %
\mbox{ steps of \textsc{AssignCode} }\\
% &   & \\
  \bullet &   0 &  \mbox{\textbf{otherwise}} \, .
\end{array}
\label{YU}
\eeq 
Denote by $\Gamma_u^{(t)}$ the set of \textit{active neighbors} of $u$ at
any given time $t$ during the execution of the algorithm.
Suppose that we are in such time slot $t$.
Independently of its previous attempts, $u$ remains uncolored with probability
\beq
p_{u,t} = {\left(1-\frac{1}{(\Delta+p_u)}\right)} + %
\frac{1}{(\Delta+p_u)} \times %
\underbrace{\left(1 - %
\left(1-\frac{1}{(\Delta+p_v)}  \right)^{|\Gamma_u^{(t)}|} \right)}_%
{\small\mbox{There is at least a collision  due %
to one neighbor $v \in \Gamma_u^{(t)}$}\normalsize} \, .
\eeq
Since $\forall t, \, |\Gamma_u^{(t)}| \leq \Delta$ 
and $\forall v, \, 1 \leq |p_v| \leq \Delta+1$,
we have 
\ben
p_{u,t} & \leq & {\left(1-\frac{1}{(\Delta+p_u)}\right)} + %
\frac{1}{(\Delta+p_u)}%
{\left(1 - %
\left(1-\frac{1}{\Delta}  \right)^{|\Gamma_u^{(t)}|} \right)} \\
  & \leq & {\left(1-\frac{1}{(\Delta+p_u)}\right)} + %
\frac{1}{(\Delta+p_u)} \times %
{\left(1 - %
\left(1-\frac{1}{\Delta}  \right)^{\Delta} \right)} \\
  & \leq & 1 - \frac{1}{e \, (\Delta + p_u)} \, \, \leq \, \, %
 1 - \frac{1}{2e\, \Delta} \, \, \leq \, \, 1 - \frac{1}{6 \Delta} \, .
\een
Therefore, with probability at most 
\beq
\left(1-\frac{1}{6 \Delta} \right)^{D_{\ell} \ln{n}^2} \leq
\exp{\left( - \frac{D_{\ell} \ln{n}^2}{6 \Delta}\right)} \, 
\eeq
$u$ remains uncolored during the whole algorithm.
Thus, the expected number of uncolored vertices at the end
of the protocol \textsc{AssignCode} is less than
\beq
\Esp{\left[Y\right]} = \sum_{u} \Esp{\left[Y_u\right]} %
\leq n \, \exp{\left( - \frac{D_{\ell} \ln{n}^2}{6 \Delta}\right)} \, .
\eeq
Since by (\ref{W0W1-bis}) we have 
\beq
\Delta = \Delta(\ell) \leq  %
- 2 \, \frac{ \ell \,\ln{n}}{W_{0}\L(- \frac{\ell}{e \, (1+\ell)}\R)} \, .
\eeq
It is now easy to choose a constant $D_{\ell}$ such that 
\beq
D_{\ell} > - \, \frac{12 \ell} {W_{0}\L(- {\ell}/{e \, (1+\ell)}\R)} \, ,
\eeq
in order to have $\Esp{\left[ Y \right]} \ll 1$ as $n \rightarrow \infty$.
After using the well known Markov's inequality (cf. \cite{ALON}),
the proof of our Theorem is now done.
\end{proof}

\section{Conclusion}
The main purpose of this paper has been that of investigating the
fundamental characteristics of a randomly deployed set of sensor
nodes. First of all, with respect to the communications, 
 the characteristics of interest include the 
diameter and the degrees (minimum and maximum) of the reachability
graph generated by the nodes. Next, taking the
 sensing range of the nodes as a parameter, several 
degrees of coverage have been characterized with rigorous proofs. 

On the top of these typical behaviors of random sensor networks,
 two distributed mechanisms 
are proposed. The first one concerns the dissemination of the identifiers
of the nodes to their neighbors while the second protocol 
 solves the code assignment problem. With high probability, both
 protocols are shown to achieve their tasks in polylogarithmic 
 time slots. 

While performance evaluations of algorithms intended for
 networks have been employed for mainly simple and regular graphs,
 our results show that current mathematical tools are available
to design and  analyze protocols intended to sensor networks.

Finally, we briefly point out a remark related to some real-life situations.
 A possible way of future investigations
could be to search for similar results as those presented here whenever 
the communication radii of the nodes do not degrade immediately but rather
 in a continuous fashion.

\bibliographystyle{plain}

\newpage

\section*{Appendix}
\begin{figure}[h]
\hfill
  \begin{minipage}[t]{6.0cm}
    \begin{center} 
      \psfig{file=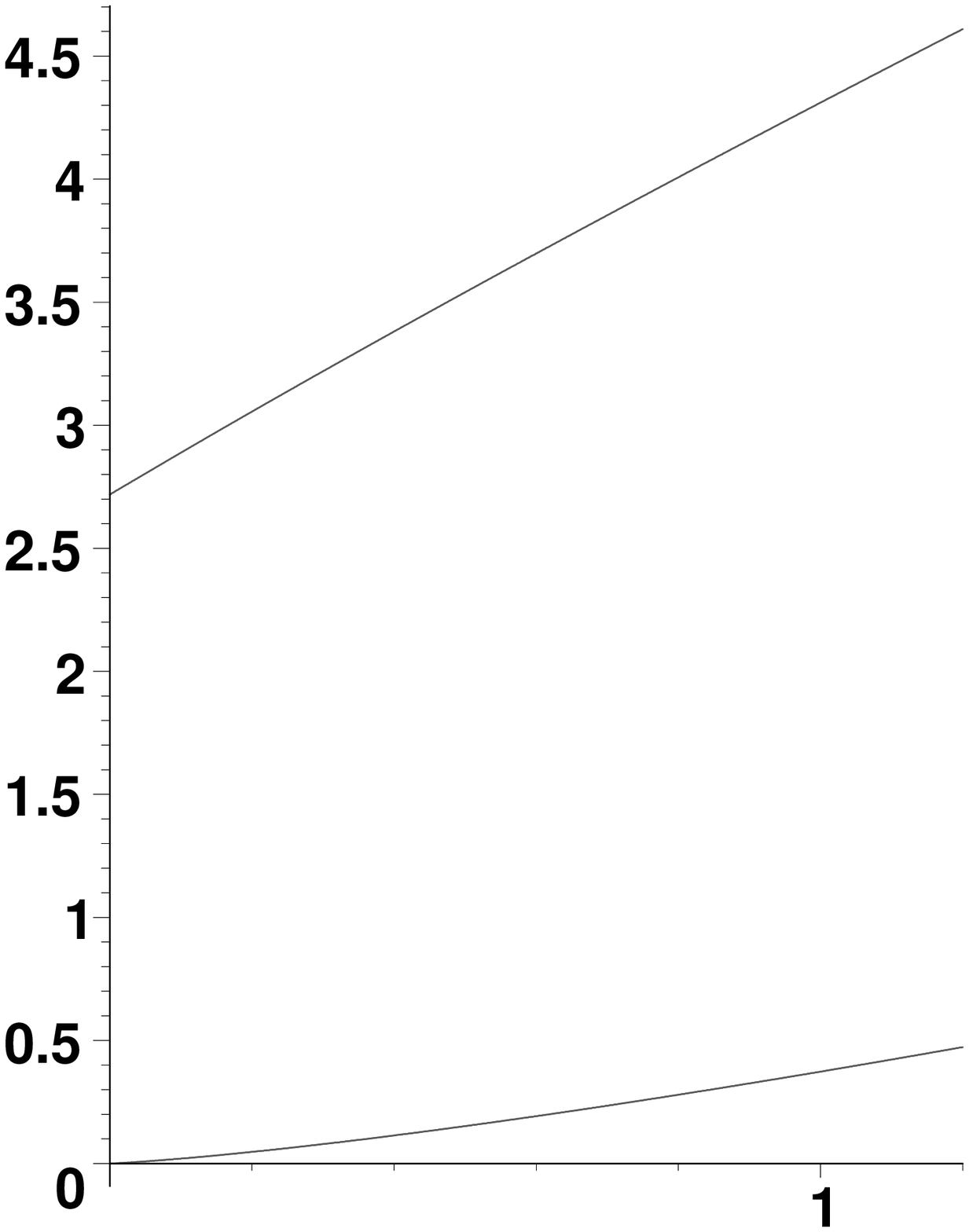,width=6.0cm,height=6.0cm}
    \end{center}
    \caption[Plots of $ -\ell/W_{-1}\L(-\frac{\ell}{e\, (1+\ell)}\R)$
and $ -\ell/W_{0}\L(-\frac{\ell}{e\, (1+\ell)}\R)$.]
    {Plots of $ -\ell/W_{-1}\L(-\frac{\ell}{e\, (1+\ell)}\R)$
and $ -\ell/W_{0}\L(-\frac{\ell}{e\, (1+\ell)}\R)$.}
    \label{FIG:PLOTW1}
  \end{minipage}
%% ICI
\hfill
  \begin{minipage}[t]{7.0cm}
    \begin{center}
      \psfig{file=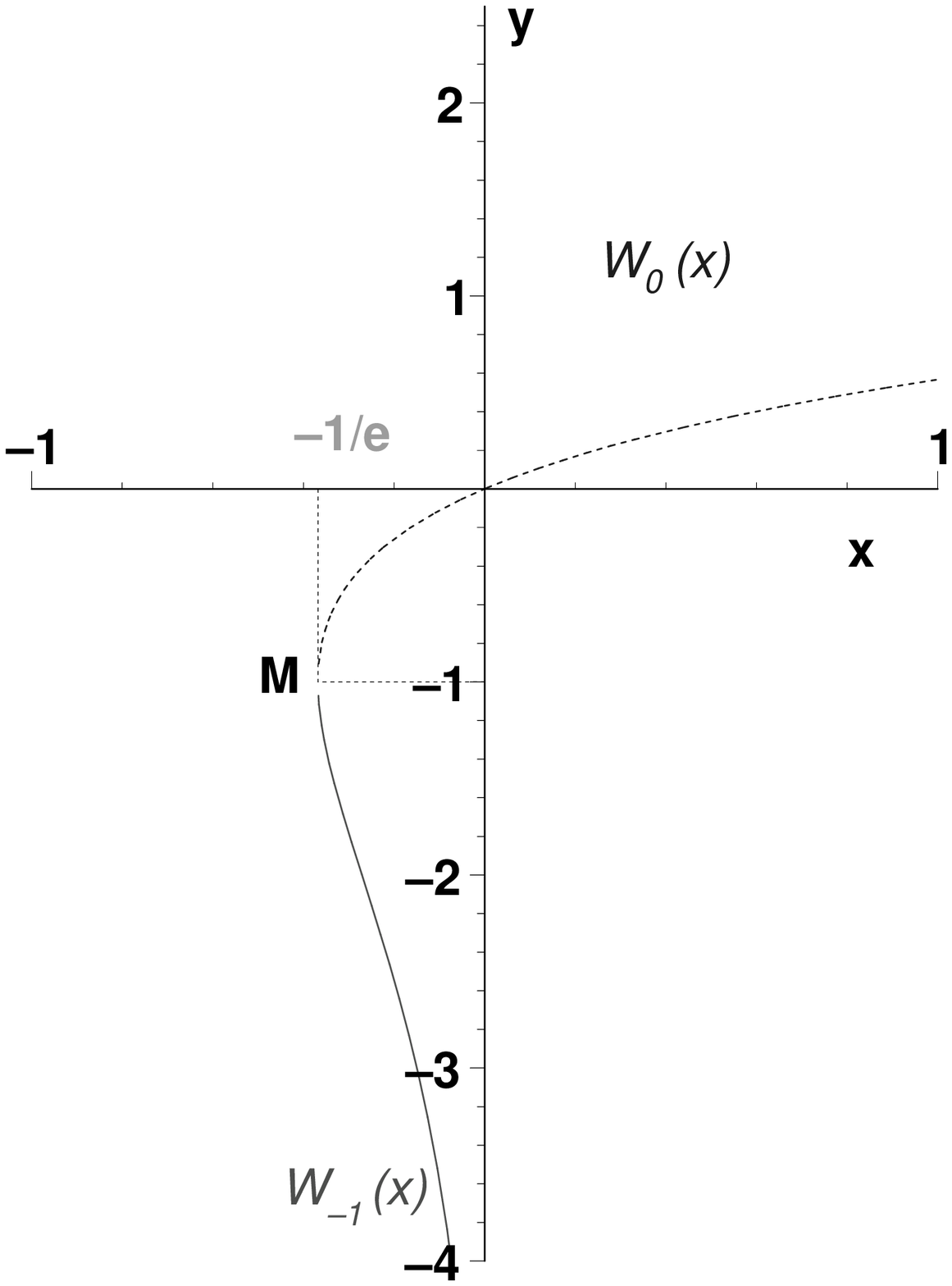,width=7.0cm,height=7.0cm}
    \end{center}
    \caption{The branches $W_0$ (dashed line) and $W_{-1}$ 
      (solid line) of the Lambert W function.}
    \label{TWO-BRANCHES}
    \end{minipage}
\hfill
\end{figure}

\noindent {\bf The Lambert W function.} In this paragraph,
we give some properties of the function satisfying $W(x)e^{W(x)} = x$.
We remark here that the function $W$, in particular
the principal branch $W_0$, already plays
a central key role when studying the random graph model
$\G(n,m)$, i.e., the random graph built with $n$ vertices and $m$ edges
 which is the ``enumerative counterpart'' of
the $\G(n,p)$ random graph model (see,  e.g., 
the ``giant paper'' \cite{JKLP93}).
In fact, $-W_0(-x)$ is the exponential generating function that enumerates
 Cayley's rooted trees \cite{CAYLEY} and we have
\beq
-W_0(-x) = \sum_{i=1}^{\infty} \frac{n^{n-1} x^n}{n!} \, .
\label{EGF-CAYLEY}
\eeq
We plot in Figure \ref{TWO-BRANCHES} the two real branches of 
the Lambert W function considered in this paper.
This function has been recognized
 as solutions of many problems in various fields of mathematics,
physics and engineering as emphasized in \cite{LAMBERTW}.
The Lambert W is considered as a special function of
mathematics on its own and its computation has been implemented
in mathematical software as Maple. Figure \ref{TWO-BRANCHES}
represents the two real branches of the Lambert W function.
It is shown that the two branches meet at point $M=(-1/e,\, -1)$.
As an example, if $\ell = \frac{1}{2}$ in (\ref{W0W1}) of Theorem \ref{MULTIPLE_COVERAGE},
each point of the area $X$ is covered, with high probability,
 by at least  $\jmin \sim .1520088850 \ln{n}$ disks. We have the Figure
\ref{FIG:PLOTW1} depicting the functions 
$\ell \ten -\ell/W_{-1}\L(-\frac{\ell}{e\, (1+\ell)}\R)$ and
$\ell \ten -\ell/W_{0}\L(-\frac{\ell}{e\, (1+\ell)}\R)$
involved in Theorem \ref{MULTIPLE_COVERAGE}.

\vspace{-0.0cm}

\end{document}